\documentclass{pasa}%

\usepackage{graphicx}

\title[The Cosmological One-Way Speed of Light]{The One-Way Speed of Light and the Milne Universe}

\author[Lewis \& Barnes]{Geraint F. Lewis$^1$\thanks{\tt geraint.lewis@sydney.edu.au} \ and Luke A. Barnes$^2$ \\
{\small $^1$Sydney Institute for Astronomy, School of Physics, A28, The University of Sydney. NSW 2006, Australia} \\
{\small $^2$Western Sydney University, Locked Bag 1797, Penrith South DC, NSW 2751, Australia}
}%

\jid{PASA}
\doi{10.1017/pas.\the\year.xxx}
\jyear{\the\year}

\usepackage{aas_macros}
\usepackage{hyperref} 
\hypersetup{colorlinks,citecolor=blue,linkcolor=blue,urlcolor=blue}

\begin{document}

\begin{frontmatter}
\maketitle

\begin{abstract}
In Einstein's Special Theory of Relativity, all observers measure the speed of light, $c$, to be the same. 
However, this refers to the round trip speed, where a clock at the origin times the outward and return trip of light reflecting off a distant mirror. Measuring the one-way speed of light is fraught with issues of clock synchronisation, and, as long as the average speed of light remains $c$, the speeds on the outward and return legs could be different. One objection to this anisotropic speed of light is that views of the distant universe would be different in different directions, especially with regards to the ages of observed objects and the smoothness of the Cosmic Microwave Background. In this paper, we explore this in the Milne universe, the limiting case of a Friedmann-Robertson-Walker universe containing no matter, radiation or dark energy. 
Given that this universe is empty, it can be mapped onto flat Minkowski space-time, and so can be explored in terms of the one-way speed of light. The conclusion is that the presence of an anisotropic speed of light leads to anisotropic time dilation effects, and hence observers in the Milne universe would be presented with an isotropic view of the distant cosmos.
\end{abstract}

\end{frontmatter}

\section{INTRODUCTION }
\label{sec:intro}
Central to Einstein's Special Theory of Relativity is that all inertial observers will measure an identical value of the speed of light, $c$ \citep{1905AnP...322..891E}. 
However, as noted by Einstein himself, this refers to the average of a round-trip journey for light that is reflected off a distant mirror, and, as long as the  average 
speed is $c$, the outward and inward velocities could be different \citep[see extensive review in][]{AndersonR1998Cosg}. 
Whilst this might seem strange, anisotropy in the speed of light would result in anisotropy in time dilation effects, ensuring that synchronisation of distant clocks remains fraught.  
Hence, no experimental measurement of the one-way speed of light is possible.

An objection to differing one-way speeds of light might be observations of the distant universe, where we clearly have, on average, an isotropic view, seeing young galaxies at high redshift, and the smoothness and uniformity of the Cosmic Microwave Background over the sky; surely the anisotropy in the speed of light would be imprinted on this view? In this paper, we will tackle this question by considering an idealised cosmological model, the Milne universe, and will explore an extreme case where the speed of light is infinite in one direction, and $c/2$ in the other.
The layout of this paper is as follows: In Section~\ref{sec:owsol} we present the mathematics of differing one-way speeds of light, and will present the Milne universe in Section~\ref{sec:milne}. We discuss the Milne universe with differing one-way speeds of light in Section~\ref{sec:mow}, whilst presenting our conclusions in Section~\ref{sec:conc}. In the following, we will set $c$, the average round trip speed of light, to unity. 

\begin{figure*}
\begin{center}
\includegraphics[width=6in]{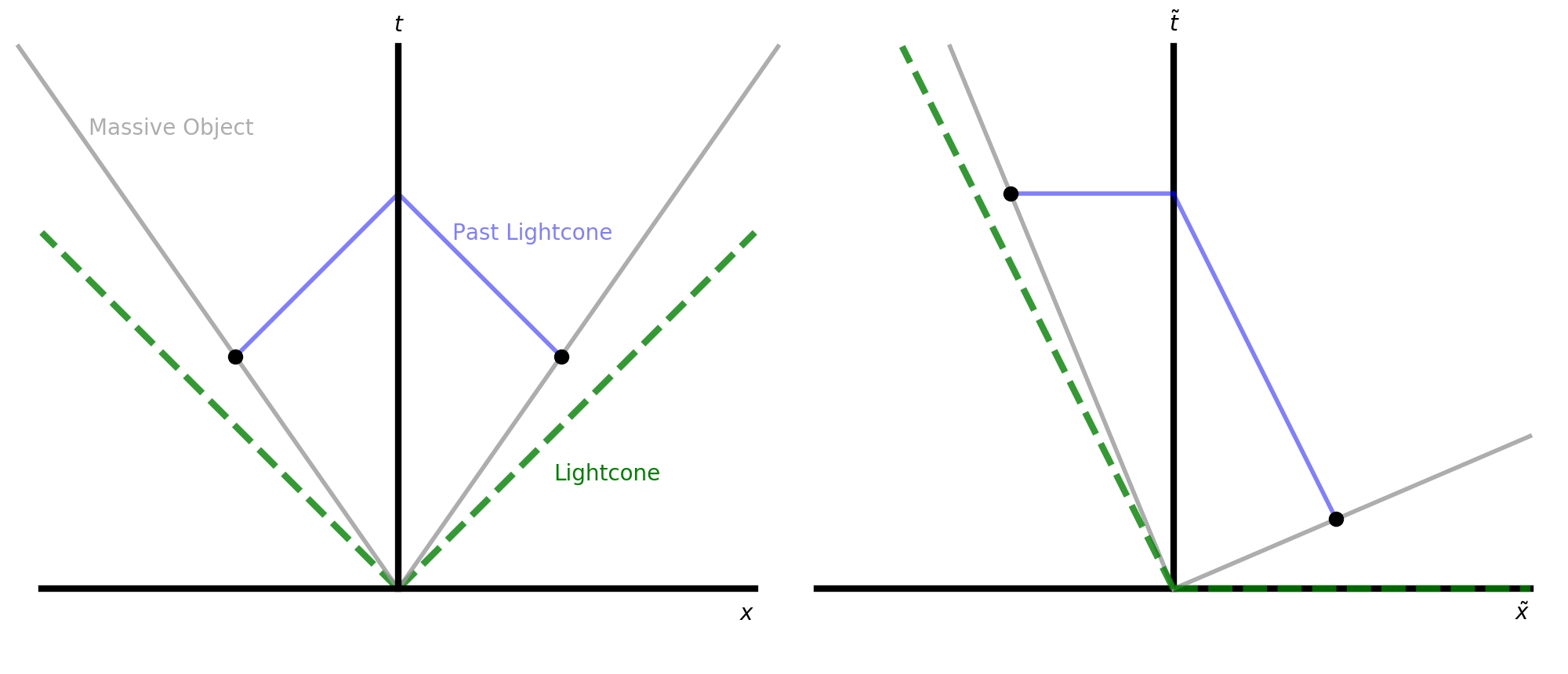}
\caption{Space-time diagram for the situation where the speed of light is equal in both directions (left) and the limiting case where the speed of light is $c/2$ in one direction, and infinite in the other (right). 
The green dashed line represents a future lightcone for the observer at the origin of space and time, whereas the grey lines represent the worldlines of massive objects moving relative to the coordinate system.
Through synchronising all clocks at the origin, 
the blue lines represent light rays  emitted from the massive objects after a fixed amount of time has passed for both. Clearly, the observer at the origin sees the massive objects at the same age when the light rays are seen.}\label{Fig1}
\end{center}
\end{figure*}

\section{ONE-WAY SPEED OF LIGHT}
\label{sec:owsol}
In considering differing one-way speed of light models, the underlying transformations of coordinates are modified. In the following, we follow the 
mathematical formalism of \citet{AndersonR1998Cosg}. We consider differing one-way speeds of light related to $c$ by
\begin{equation}
    c_\pm = \frac{ c }{ 1 \mp \kappa }  \equiv \frac{ 1 }{ 1 \mp \kappa }
    \label{eqn:speedlight}
\end{equation}
Setting $\kappa=0$ corresponds to an isotropic speed of light, whereas $\kappa=1$ presents the extreme case where the $c_+ = \infty$ and $c_- = 1/2$. To preserve the observations of special relativity, a coordinate velocity $v = \frac{dx}{dt}$ in the isotropic $c$ case ($\kappa=0$) is mapped to a new velocity 
\begin{equation}
    \tilde{v} = \frac{d\tilde{x}}{d\tilde{t}}  = \frac{\gamma}{\tilde{\gamma}}v
    \label{eqn:vel}
\end{equation}
where
\begin{equation}
    \gamma = \frac{1}{\sqrt{ 1 - v^2 }}
    \ \ \
    {\rm and}
    \ \ \ 
    \tilde{\gamma} = \frac{ 1 - {\kappa v} }{\sqrt{ 1 -  v^2  }}
    \label{eqn:gamma}
\end{equation}
The relative time dilation between two observers in the case where the one-way speed of light is given by
\begin{equation}
    \frac{d\tilde{t}'}{d\tilde{t}} = \frac{1}{\tilde{\gamma}}
    \label{eqn:timedilation}
\end{equation}

Figure~\ref{Fig1} presents an illustration of the impact of the anisotropic speed of light. The left-hand side of this figure presents the familiar case where the speed of light is equal in both directions with the green-dashed line representing a light cone for an observer at the origin. The grey lines represent the worldlines of massive objects moving relative to the observer at the origin at $v = 0.7$, with their clocks synchronised at the origin. The  blue lines represent light rays emitted from massive objects after a fixed about of proper time has passed. Given the symmetry of the situation, the light rays arrive at the origin at the same time, and the observer sees the moving objects showing the same amount of proper time passing at the same instant.

\section{THE MILNE UNIVERSE}
\label{sec:milne}
The discussion in the previous section is wrapped in the language of special relativity, whereas the cosmological description of the universe relies on Einstein's General Theory of Relativity. Whilst the typical approach to studying cosmology is to begin with the Friedmann-Robertson-Walker (FRW) metric \citep[e.g.][]{hobson_efstathiou_lasenby_2006}, this is just a convenient choice of coordinates and other choices can be made. \citet{1945PhRv...68..250I} demonstrated that cosmological models can be cast in a kinematic form, where by motion through space, coupled with gravitational potentials, replaces the picture of expanding space  \citep[e.g.][]{2007MNRAS.381L..50L}.

The focus of this paper will be the limiting case of the FRW metric, namely the Milne universe in which the universe is empty, devoid of any matter, radiation or energy \citep{1933ZA......6....1M}. A key feature of of the Milne universe is that, while it is spatially curved, its space-time is flat and can, therefore, be directly mapped into the Minkowski metric \citep[see][]{2005PASA...22..287C}

\begin{figure*}
\begin{center}
\includegraphics[width=6in]{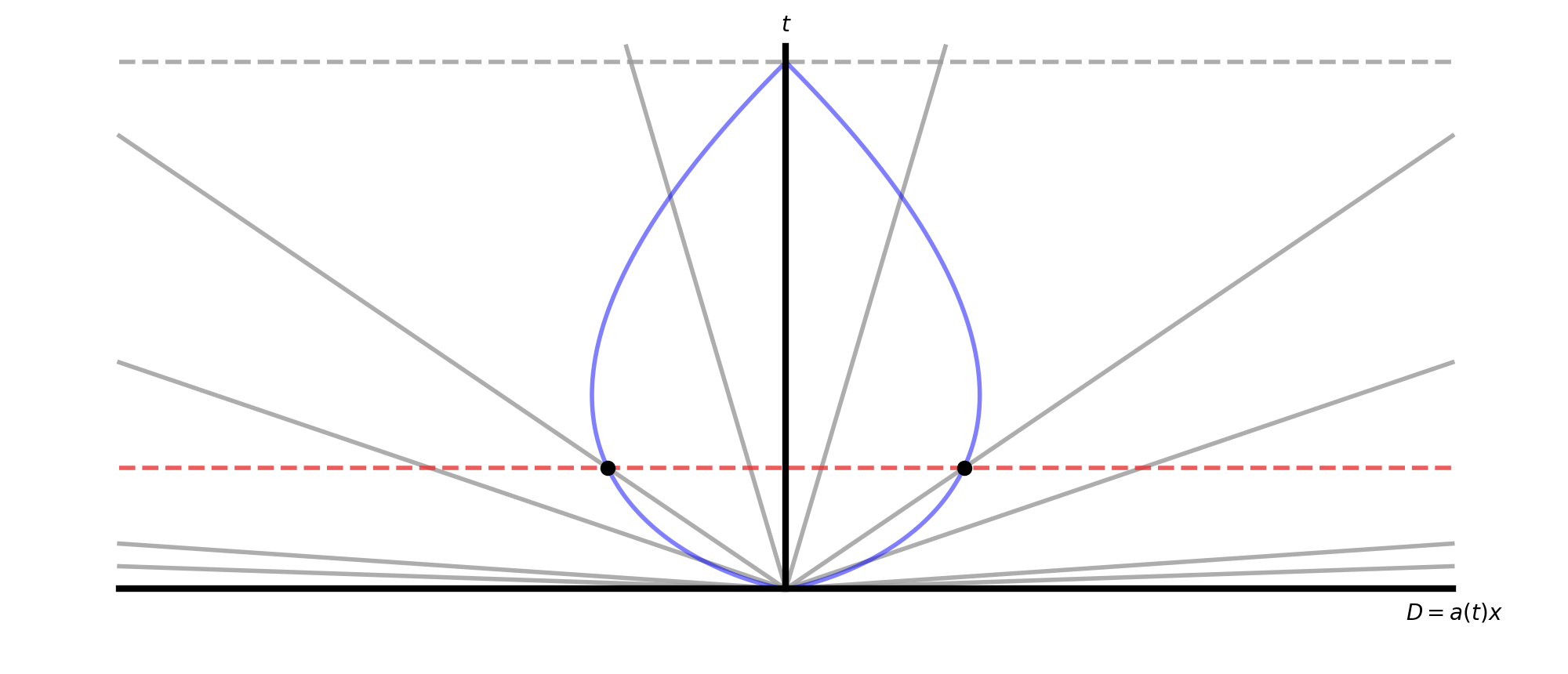}
\caption{Space-time diagram for the Milne universe in FRW coordinates. The horizontal dashed grey line denotes now in cosmic time, whilst the sold grey lines are comoving objects at $x = 1, 5, 10, 50$ and $100$. The blue lines represent the past light cone for an observer at the origin today and the time where they cross the comoving objects is the age we observe them at today; clearly, due to the symmetry of the situation, the view in opposite directions will be the same, with more distant objects appearing younger.
The two black dots denote emission from $x=5$ that is observed at the origin today, whilst the red dashed line represents the age of the universe when the light from these sources is emitted.}\label{Fig2}
\end{center}
\end{figure*}


We begin by describing the Milne universe in Minkowski space-time. At $t = 0$, a collection of massive test particles (quaintly referred to as \emph{galaxies}) are ejected in all directions from the origin ($x = 0$) with a range of velocities. In the limiting case of an empty universe, we disregard gravity (the effect of these galaxies on space-time), and so the galaxies maintain a constant velocity. Because faster galaxies move further in a given period of time, the further away we look, the faster the galaxies are moving and the more their light is redshifted; this gives the Hubble law for any of the galaxies.

Assuming an isotropic speed of light, consider two galaxies that are emitted with the same speed in opposite directions. In order to measure their positions, we (still at the origin) send a light beam after them at time $t_1$. The light bounces off the galaxy at $(x_g,t_g)$ and returns to us at $t_2$. With an isotropic speed of light, the light reached the galaxy halfway between $t_1$ and $t_2$. The distance to the galaxy is half of the total light travel time: 
\begin{align}
t_g &= \frac{1}{2} (t_2 + t_1) \\
x_g &= \frac{1}{2} (t_2 - t_1) ~.
\end{align}
This second expression effectively defines the radar distance to an object \citep[c.f.][]{2008MNRAS.388..960L}.
If the galaxy started a clock as it departed the origin, the time on that clock when our photon arrives is given by,
\begin{equation}
\tau_p = \sqrt{x_g^2 - t_g^2} = \sqrt{t_1 ~ t_2} ~.
\end{equation}
From this, we infer that the galaxy is moving with speed $v_g = x_g/t_g = (t_2 - t_1) / (t_2 + t_1)$.

Now, suppose that the speed of light is $c_+ = \infty$ in the positive $x$-direction and $c_- = 1/2$ in the negative $x$-direction, as shown in Figure \ref{Fig1} (right). As before, we send a beam of light after the galaxy at $t_1$ and it returns to us at $t_2$.

\paragraph{Right-hand Galaxy (moving in the positive direction):} The light travels instantaneously to the galaxy at $t_1$, and returns to us at speed $1/2$,
\begin{align}
t_{gr} &= t_1 \\
x_{gr} &= \frac{1}{2} (t_2 - t_1) ~.
\end{align}
For this anisotropic universe, the formula to calculate the proper time is,
\begin{equation}
\tau_{gr} = \sqrt{t_{gr}^2 + 2 x_{gr}t_{gr}} = \sqrt{t_1 ~ t_2} ~,
\end{equation}
as above. The inferred speed of the galaxy is $v_{gr} = x_{gr}/t_{gr} = -\frac{1}{2}(t_2 - t_1) / t_1$, which is related to the inferred velocity for the isotropic case by Equation \eqref{eqn:vel}.

\paragraph{Left-hand Galaxy (moving in the negative direction):} The light travels at speed $c/2$ to the galaxy at $t_1$, and returns instantaneously to us,
\begin{align}
t_{gl} &= t_2 \\
x_{gl} &= -\frac{1}{2} (t_2 - t_1) ~.
\end{align}
For this anisotropic universe, the formula to calculate the proper time is,
\begin{equation}
\tau_{gl} = \sqrt{t_{gl}^2 + 2 x_{gl}t_{gl}} = \sqrt{t_1 ~ t_2} ~,
\end{equation}
as above. The inferred speed of the galaxy is $v_{gl} = 
|x_{gl}|/t_{gl} = \frac{1}{2}(t_2 - t_1) / t_2$, which is related to the inferred velocity for the isotropic case by Equation \eqref{eqn:vel}.

Importantly, the redshift of the returning light that is observed by us is the same in all three cases: $1 + z_g = \sqrt{t_2/t_1}$, so the observed universe is identical whether the speed of light is identical in all directions, or is anisotropic.
However, for the anisotropic  advocate, galaxies with the same redshift in different directions are located at the same distance, but emitted the light we observe at different times. They have different velocities, and so they conclude that the left side of the universe is expanding faster than the right hand side.


\begin{figure*}
\begin{center}
\includegraphics[width=6in]{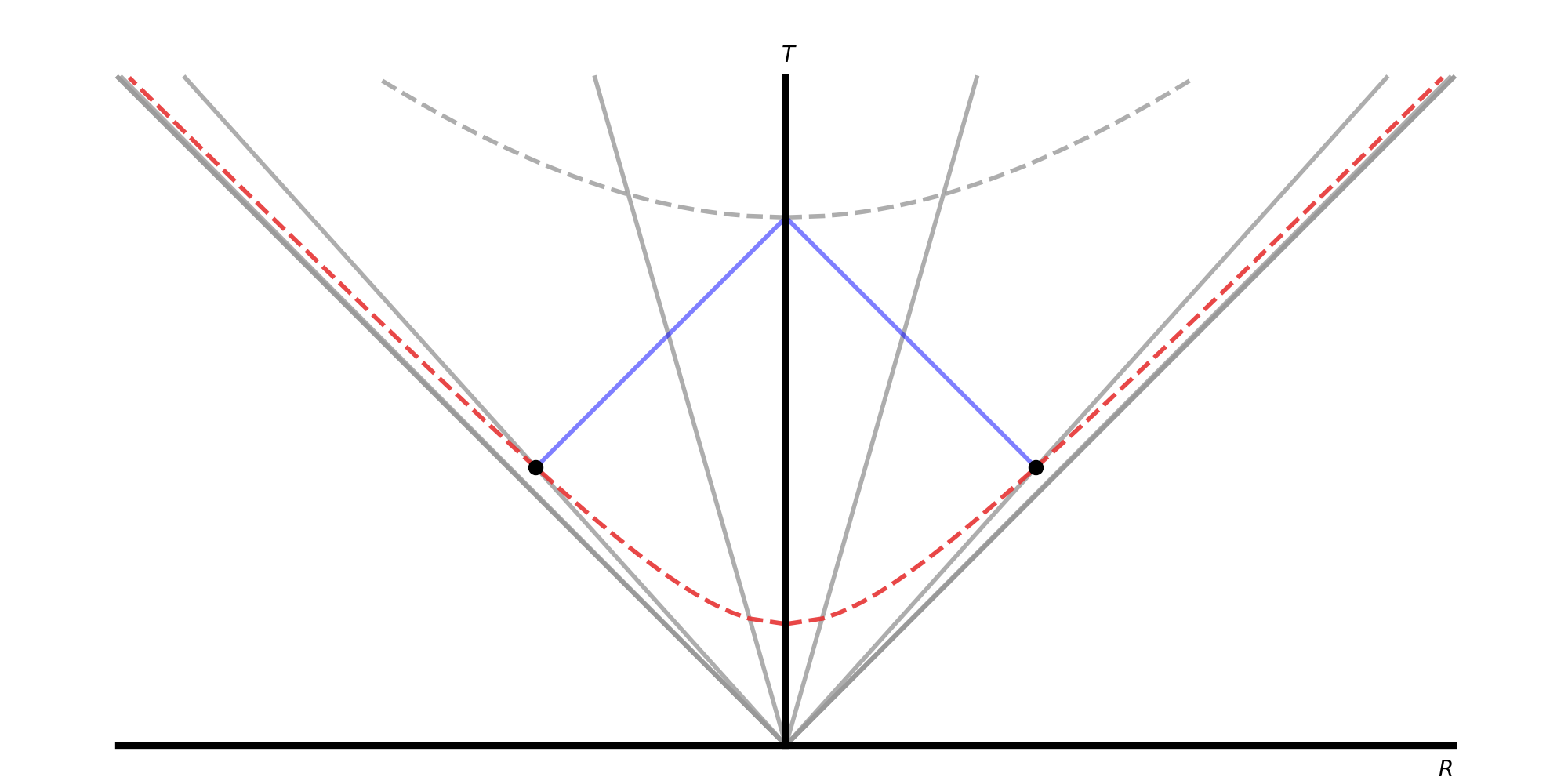}
\caption{The Milne universe presented in Figure~\ref{Fig2}, but now mapped into the flat space-time coordinates. The comoving objects have been mapped into sloped lines, whereas synchronised lines of constant cosmological time have been mapped into hyperbola. There are clear similarities between this and the left-hand space-time diagram presented in Figure~\ref{Fig1}.
}\label{Fig3}
\end{center}
\end{figure*}

What does this Milne universe look like in an expanding space framework? We begin with the the FRW metric,
\begin{equation}
    ds^2 = -dt^2 + a^2(t) \left[ dx^2 + R_o^2 S_k^2\left(x/R_o\right) d\Omega^2 \right]
    \label{eqn:FRW}
\end{equation}
where $a(t)$ is the normalised scale factor, such that $a(t_o)=1$ and $t_o$ is the present age of the universe. The function, $S_k(x)$ is $\sin(x)$, $x$, and $\sinh(x)$ for a spatially closed, flat and open universe respectively. The angular terms, which are related to the surface of a 3-sphere, are given by $d\Omega^2 = d\theta^2 + \sin^2\theta\ d\phi^2$. The present day scale factor, $R_o$, in an open universe is given by
\begin{equation}
    R_o = \frac{1}{H_o} \frac{1}{\sqrt{ 1 - \Omega_o }}
    \label{eqn:scale}
\end{equation}
where $H_o$ is the present day Hubble Constant and $\Omega_o$ is the present day total energy density. 
For the Milne universe, $\Omega_o = 0$, and so $R_o = H_o^{-1}$, and the normalised scale factor $a( t ) = t / t_o$.

It is instructive to construct a space-time diagram for the Milne universe (Figure~\ref{Fig2}) which shows the instantaneous proper distance to a comoving observer at a spatial coordinate, $x$, given by $D(t) = a(t) x$ verses the cosmological time, $t$; as an illustration, comoving observers are presented at $x  = 1, 5, 10, 50$ and $100$. Also, presented in blue, is the past lightcone for an observer at the spatial origin at the present time. The path of a light ray in these coordinates is governed by
\begin{equation}
    \frac{dx}{dt} = \pm \frac{t_o}{t}
    \label{eqn:slope}
\end{equation}
Remembering that in these coordinates, the proper times of comoving observers are synchronised with the cosmic time, $t$, the observer at the origin will see distant objects with an age given by their crossing of the past light cone, and, given the symmetry of the situation, the origin observers view will be symmetrical, with more distant objects appearing younger.

\section{MILNE AND SPECIAL RELATIVITY}
\label{sec:mow}
Examining Figure~\ref{Fig2} suggests that the Milne universe is very different to the flat space-time of special relativity. 
However, given that it has no material content, the underlying space-time structure of the two are the same, and so we should be able to undertake a coordinate transformation between the two. Note that this is different to the conformal representation of FRW universes \citep[e.g.][]{1991ApJ...383...60H}, which straightens light rays to $45^o$, as there can be a complex relationship between universal conformal time and the experienced proper time.

We follow \citet{2007MNRAS.378..239C} by firstly defining $\chi = x / R_o$ and $dt = R_o a d\eta$ such that Equation~\ref{eqn:FRW} can be written as
\begin{equation}
    ds^2 = R_o^2 a^2(\eta) \left[ -d\eta^2 + d\chi^2 + \sinh^2( \chi ) d\Omega^2 \right]
    \label{eqn:FRW_new}
\end{equation}
\begin{figure*}
\begin{center}
\includegraphics[width=6in]{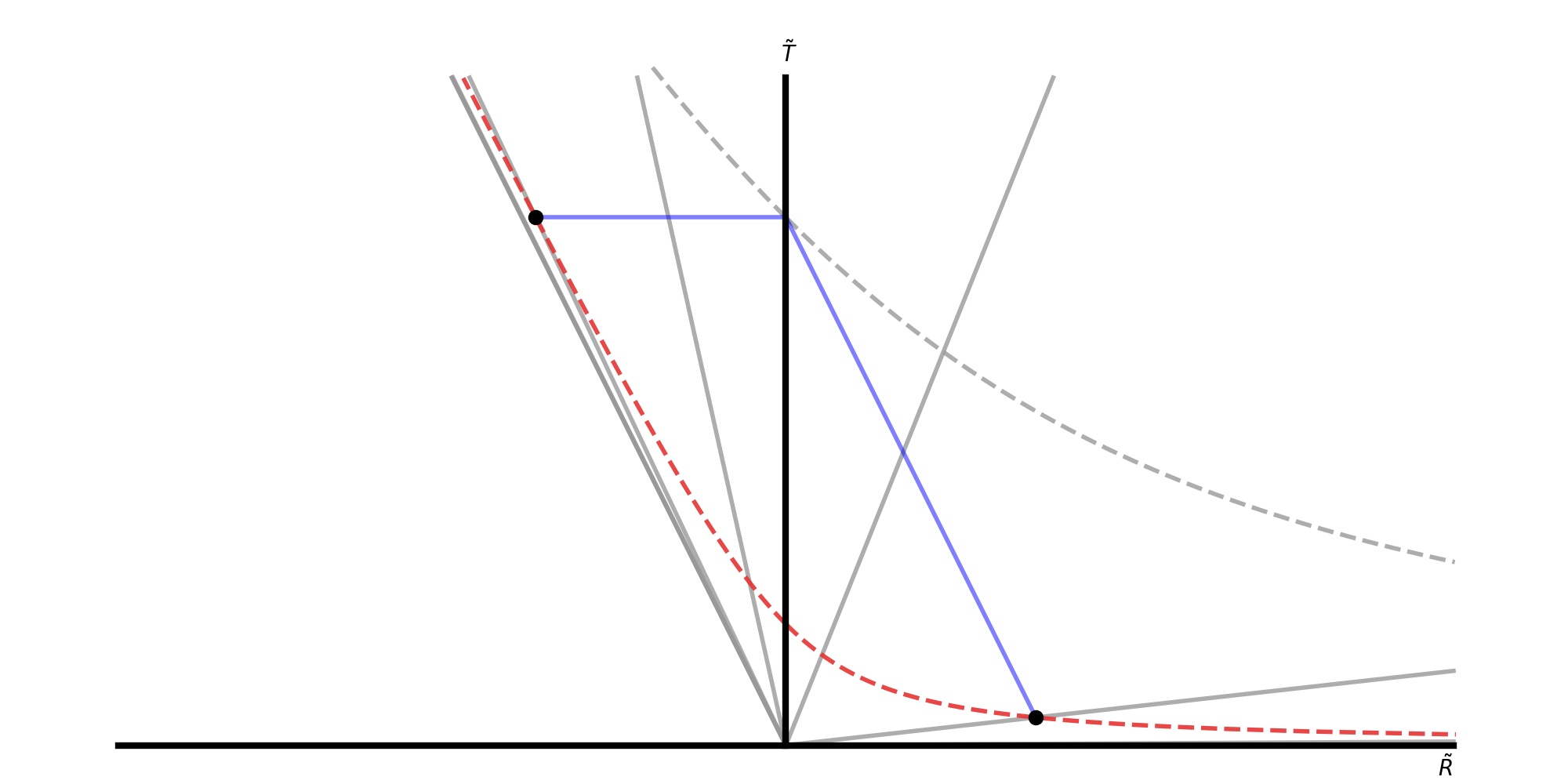}
\caption{The Milne universe presented in Figure~\ref{Fig2}, but mapped into the case with an anisotropic speed of light. Again, the grey dashed line corresponds to the present time in the cosmological time of the Milne universe, whereas the the red dashed line is the cosmological time for a pair of emitters on either side of the sky. As can be seen, the observer at the origin is presented with an isotropic view of the sky, even through the speed of slight is anisotropic.
}\label{Fig4}
\end{center}
\end{figure*}
At this stage, we define a coordinate transformation from $(\eta,\chi)$ to new coordinate $(T,R)$ through
\begin{eqnarray}
T = A e^\eta \cosh(\chi) \\
R = A e^\eta \sinh(\chi)
\end{eqnarray}
where $A$ is a constant. With these
\begin{equation}
    A e^\eta = \sqrt{ T^2 - R^2 }\ \ \ {\rm and}\ \ \ \tanh(\chi) = \frac{R}{T}
    \label{eqn:transform}
\end{equation}
with these transformations, and a little algebra, Equation~\ref{eqn:FRW_new} can be written as
\begin{equation}
    ds^2 = -dT^2 + dR^2 + R^2 d\Omega^2
    \label{eqn:flat}
\end{equation}
which is just the flat space-time of special relativity with polar coordinates over the spatial part. Clearly, in these new coordinates, light rays travel at $45^o$, comoving observers are represented as straight lines with a slope given by $\tanh(\chi)$. Additionally, the relationship between the proper time experienced by an observer at the origin, $dT$, and the comoving observer, $d\tau$, is given by
\begin{equation}
    \frac{d\tau}{dT} = \sqrt{ 1 - \left( \frac{dR}{dT} \right)^2 } = \sqrt{ 1 -  \tanh^2(\chi)  }
\end{equation}

To illustrate this, we map the situation presented in Figure~\ref{Fig2} into the $(T,R)$ coordinates. This is illustrated in Figure~\ref{Fig3}. As noted above, the comoving observers to sloped lines, and as can be seen, these asymptote to $45^o$ as all motion is bounded by the speed of light. In these coordinates, the comoving observers in the Milne universe, which are moving apart due to expanding space, are transformed into objects moving with velocities relative to each other through flat space-time. The key points that synchronous lines in the Milne universe, which are lines of constant universal time, have been mapped into hyperbola in the $(T,R)$ coordinates.  

The remaining question is: how do lines of constant cosmological time map onto the situation where the one-way speed of light is anisotropic? Hence we undertake the transformation of the situation in Figure~\ref{Fig3} through the mathematics present in Section~\ref{sec:owsol}, noting that the velocity is given by $v = \frac{dR}{dT}$, and present the result in Figure~\ref{Fig4}. Again, the extreme case is considered, so the speed of light in one direction is $1/2$ and the other is infinite.

\begin{figure*}
\begin{center}
\includegraphics[width=6in]{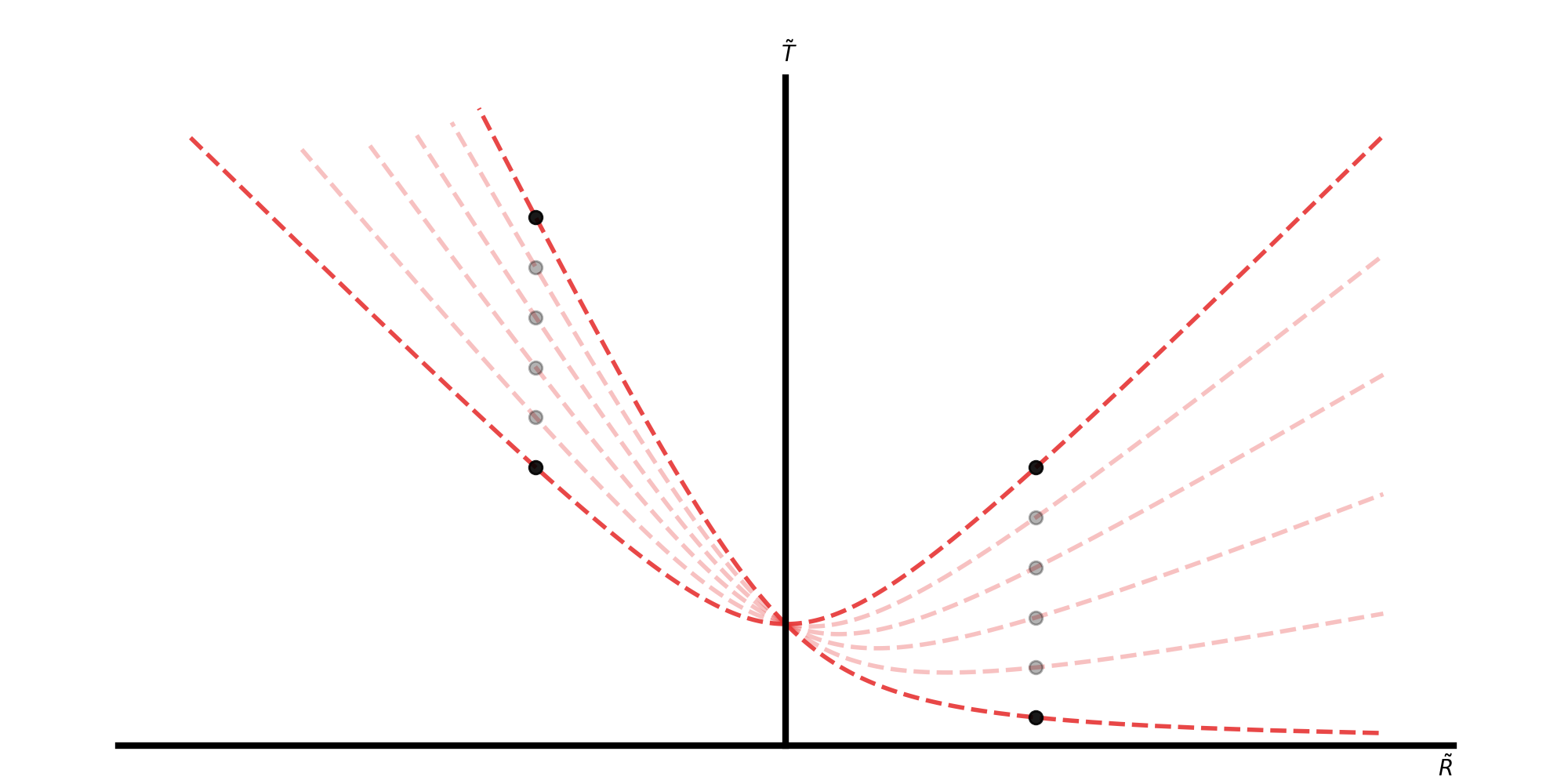}
\caption{The line of simultaneity in the emitter in the FRW coordinates (Figure~\ref{Fig2}), mapped into the anisotropic velocity of light coordinates (Section~\ref{sec:mow}) from $\kappa=0$, the isotropic case represented as a hyperbola, to the extreme case, with $\kappa=1$, both presented in bolder red, with intermediate cases, in steps of $\kappa=0.2$, presented in lighter red. The filled circles represent the location of the emitter in these coordinates for each of the cases. Note that the spatial location of the emitter in the $\tilde{R}$ coordinate is independent of $\kappa$.}\label{Fig5}
\end{center}
\end{figure*}

As seen in Figure~\ref{Fig1}, the comoving observers are now asymmetric about the observer at the origin, but also added to Figure~\ref{Fig4} are lines of synchronicity of the FRW universe, with the grey dashed line being now, whereas the red dashed line is the time that two sources emit their light to be observed by the observer at the origin (see Figure~\ref{Fig2}). As can be seen, these are now asymmetric due to the asymmetric action of time dilation. In fact, the difference in the one-way speed of light and the resultant time dilation conspire to ensure that the view of the observer at the origin is identical irrespective of the one-way speed of light. Hence, at least in the Milne universe, whether the speed of light is identical in all directions, or is anisotropic, results in a uniform view of the distant universe.

So far, we have considered two cases, where either the speed of light is isotropic, or the extreme case where the anisotropic speed is $1/2$ in one direction, and infinite in the other. The question remains whether this holds true in general case, for an arbitrary $\kappa$. To consider this, we explore the form of the synchronous lines in the FRW universe (lines of constant cosmological time) in the Milne universe when we consider the anisotropic speed of light. 
In Figure~\ref{Fig5} the emitters from the previous discussion are presented as filled circles, but now values of $\kappa$ of $0$ to $1$ in steps of $0.2$ are considered; as can be seen, as $\kappa$ increases, the shape of this synchronous lines becomes more asymmetrical. Also shown, as filled circles, is the location of the emitters for each of the considered values of $\kappa$; importantly, it should be noted that their location in $\tilde{R}$ is fixed.

We can explore the impact of differing values of $\kappa$ in terms of geometry in the space-time diagram. Consider an observer located at $\tilde{T} = \tilde{t}_o$ at the origin, and emitters who have experienced a proper-time, $\tilde{\tau}_e$ since leaving the origin. For an arbitrary value of $\kappa$, and noting that the location of the emitter when a photon is emitted, $(\tilde{t}_r,\tilde{r}_e)$, is given by
\begin{equation}
    \tilde{r}_e = \tilde{v}\ \tilde{t}_e
    \label{eqn:location}
\end{equation}
and noting that, from Equation~\ref{eqn:timedilation}, that
\begin{equation}
    \tilde{t}_e = \frac{ 1 - k v }{\sqrt{ 1 - v^2 }}\ \tilde{\tau}_e
    \label{eqn:emittime}
\end{equation}
then, using the definition of the velocity in these coordinates (Equations~\ref{eqn:vel}),
\begin{equation}
    \tilde{r}_e = \frac{v}{\sqrt{ 1 - v^2 }}\ \tilde{\tau}_e
    \label{eqn:re}
\end{equation}
which is independent of $\kappa$, as expected \citep[for more explanation, see the detailed discussion of the transformations presented in][]{AndersonR1998Cosg}. Hence, the speed of a light ray connecting the emitter and the observer is given by
\begin{equation}
    \frac{d\tilde{r}}{d\tilde{t}} = \frac{\Delta \tilde{r}}{\Delta \tilde{t}} = \frac{ \tilde{r}_e }{ \tilde{t}_o - \tilde{t}_e} 
    \label{eqn:speed}
\end{equation}
Noting that this speed equal $\pm 1 $ for the case where $\kappa=0$, this expression becomes
\begin{equation}
    \frac{d\tilde{r}}{d\tilde{t}} = \frac{ -sgn( v ) }{ 1 + sgn( v ) \kappa}
    \label{eqn:speed2}
\end{equation}
where $sgn( v )$ is the sign of the velocity. When $\kappa=1$, this recovers the velocity of light as being $1/2$ and infinite (c.f. Equation~\ref{eqn:speedlight}), as expected, but also illustrates that for all other values of $\kappa$, the observer will see an isotropic universe.

\section{CONCLUSION}
\label{sec:conc}
In his formulation of the special theory of relativity, Einstein chose the convention that the speed of light is isotropic and so equal in all directions. 
He also acknowledged that the physical predictions of his theory will be unchanged if the speed of light was anisotropic, as long as the average round-trip speed is equal to $c$. In this paper, we have considered the question of the impact of the one-way speed of light on cosmological observations, addressing the suggestion we should observe different sides of the sky possessing different ages if light speed was unequal. By examining the simplest FRW universe, namely the empty Milne universe, it is seen that the anisotropic speed of light results in anisotropic time dilation effects that compensate for the differing light travel times. In this universe, any observer would see an isotropic universe around them, even if the speed of light was not.

The anisotropic speed of light advocate must conclude that galaxies that are a given distance away have a faster recession speed in one direction than in the other, and the universe is expanding faster to the right or to the left. However, the dependence of redshift on the speed of light means that this does not change the appearance of the night sky. One side of the sky is not significantly more redshifted than the other. The initial velocities given to the galaxies at the beginning of the universe had just the right degree of anisotropy to balance the effect of anisotropic redshift. Perhaps at this point, the anisotropic speed of light advocate will suspect a fix, but there is nothing internally inconsistent or in contradiction with data about this model.

Of course, the Milne universe, as a limiting case of FRW cosmologies, is special in that it can be directly mapped onto the flat space-time of special relativity, and hence the question of the one-way speed of light can be directly addressed. For more general cosmological models, where the presence of mass and energy results in curved space-time, the picture is more complicated as there is no simple mapping of the modified Lorentz transformations into the general relativistic picture. We leave this discussion for a future contribution.

\begin{acknowledgements}
We thank the anonymous referee for their positive comments on this paper.
GFL thanks Derek Muller (@veritasium) and Petr Lebedev for discussions on the one-way speed of light that sparked this current study. These were part of a consultation that led to the
preparation of an informative YouTube presentation on the topic (https://youtu.be/pTn6Ewhb27k).
\end{acknowledgements}

\bibliographystyle{pasa-mnras}
\bibliography{milne}

\end{document}